\documentclass[aps,prd,showpacs,preprintnumbers,twocolumn]{revtex4}
\usepackage{graphicx}
\usepackage{epsfig}
\usepackage{amsmath}
\usepackage{epsf,amssymb,amsfonts,amsmath}

\newcommand{\beqn}{\begin{eqnarray}}
\newcommand{\eeqn}{\end{eqnarray}}
\newcommand{\eq}[1]{(\ref{#1})}
\newcommand{\be}{\begin{equation}}

\newcommand{\ee}{\end{equation}}
\newcommand{\bea}{\begin{eqnarray}}
\newcommand{\eea}{\end{eqnarray}}

\newcommand{\ep}{\epsilon}
 \usepackage{bm}
\begin{document}

\title{Charged and neutral vector meson under magnetic field}

\author{Hao Liu$^{1}$}
\email{haoliu@mail.ihep.ac.cn}
\author{Lang Yu$^{1}$}
\email{yulang@mail.ihep.ac.cn}
\author{Mei Huang$^{1,2}$}
\email{huangm@mail.ihep.ac.cn}
\affiliation{$^1$ Institute of High Energy Physics, Chinese Academy of Sciences,
Beijing 100049, China}
\affiliation{$^2$ Theoretical Physics Center for Science Facilities,
Chinese Academy of Sciences,
Beijing 100049, China}

\begin{abstract}
The vector meson $\rho$ in the presence of external magnetic field has been investigated
in the framework of the Nambu--Jona-Lasinio model, where mesons are constructed by
infinite sum of quark-loop chains by using random phase approximation.
The $\rho$ meson polarization function is calculated to the leading order of $1/N_c$
expansion. It is found that the constituent quark mass increases with magnetic field,
the masses of the neutral vector meson $\rho^{0}$ with spin component $s_z=0,\,\pm1$
and the charged vector meson $\rho^{\pm}$ with $s_z=0$ also increases with magnetic field.
However, the mass square of the charged vector meson $\rho^{+}$ ($\rho^{-}$) with
$s_z=+1$ ($s_z=-1$) decreases linearly with magnetic field and drops to zero at the
critical magnetic field $e B_c \simeq 0.2 {\rm GeV}^2$, which indicates the possible
condensation of charged vector meson in the vacuum. This critical magnetic field is
much lower than the value $eB_c=0.6 {\rm GeV}^2$ predicted by a point-like vector
meson. We also show that if we use lowest Landau level approximation, the mass of the
charged vector meson $\rho^{\pm}$ for $s_z=\pm1$ cannot drop to zero at high magnetic fields.
\end{abstract}
\pacs{12.38.-t, 13.40.-f, 25.75.-q}
\maketitle

\section{Introduction}

Strong magnetic fields with strength of $10^{18}\sim 10^{20} \textmd{G}$
(corresponding to $eB\sim (0.1-1.0~{\rm GeV})^2$), can be generated in the laboratory
through non-central heavy ion collisions~\cite{Skokov:2009qp,Deng:2012pc} at the
Relativistic Heavy Ion Collider (RHIC) and the Large Hadron Collider (LHC). In the
surface of magnetars, magnetic fields can reach $10^{14-15} G$, which is thousand of
times larger than that of an average pulsar, and in the inner core of magnetars
the magnetic fields could reach as high as $10^{18}\sim 10^{20}\textmd{G}$. Therefore,
it is important to understand the properties of Quantum Chromodynamics (QCD) vacuum and
hot/dense quark matter under strong magnetic fields. Many progresses
have been made in this field, for example, the Chiral Magnetic Effect (CME)
\cite{Kharzeev:2007tn,Kharzeev:2007jp,Fukushima:2008xe}, Chiral Vortical Effect (CVE)
\cite{KharzeevSon:2010gr}, the Magnetic Catalysis and Inverse Magnetic Catalysis \cite{Bali:20111213}, and the Vacuum Superconductor \cite{Chernodub:2010qx,Chernodub:2011mc}.
For reviews please refer to Refs. \cite{Kharzeev-re:2012ph,Liao-rw:2014ava}.

The heavy ion collision experiment provides a unique environment to search for the possibility
of local parity violation and anomalous transport effects.
The QCD vacuum has a non-trivial topological structure, at high temperatures, because sphaleron
transitions between distinct classical vacua cause an imbalance between the number of quarks
with different chirality, and results in a violation of the $\mathcal{P}$- and
$\mathcal{CP}$-symmetry. In the presence of a strong magnetic field, an electromagnetic
current can be generated along the magnetic field, which is called the anomalous Chiral Magnetic Effect (CME) \cite{Kharzeev:2007tn,Kharzeev:2007jp,Fukushima:2008xe}, and
will induce the Charge Separation Effect. Recently, the observation of charge azimuthal correlations~\cite{Abelev:2009ad,Abelev:2012pa} from RHIC and LHC possibly resulting from
the anomalous Chiral Magnetic Effect (CME).

Chiral symmetry breaking and restoration under strong magnetic field has been discussed for
many years. It has been recognized since 1990s' that the chiral condensate increases thus
the critical temperature of chiral phase transition should also increase with magnetic field,
which is the well-known magnetic catalysis \cite{Klevansky:1989vi,Klimenko:1990rh,Gusynin:1995nb}.
However, the lattice group \cite{Bali:20111213} observed the inverse magnetic catalysis
around $T_c$, which turns out to be a surprise and puzzle. There have been several proposals~\cite{Fukushima:2012kc,Kojo:2012js, Bruckmann:2013oba,Chao:2013qpa,Yu:2014sla}
trying to understand the underlying mechanism of inverse magnetic catalysis.
It is proposed that the chirality imbalance induced by sphaleron transition \cite{Chao:2013qpa}
or instanton--anti-instanton molecule pairing \cite{Yu:2014sla} can naturally explain
the inverse magnetic catalysis near $T_c$.

The vacuum superconductor was firstly proposed in hadronic models \cite{Chernodub:2010qx}
based on the energy of a free particle under magnetic fields by neglecting the internal
structure of vector mesons. More calculations have been done by using different models
in Refs.\cite{Hidaka:2012mz,Chernodub:2012zx,Chernodub:2013uja,Li:2013aa} and Refs.\cite{Frasca:2013kka,Andreichikov:2013zba,Wangkunlun:2013}, however,
whether there exists vacuum superconductor is still under debate.

Considering a free charged relativistic particle with mass $m$, electric charge $q$ and
spin $s$, moving in a homogeneous background of an external magnetic field $B$ directed
along the $z$ axis, the relativistic energy levels $\varepsilon$ of this particle are
given by the following formula:
\beqn
\varepsilon_{n,s_z}^2(p_z) = p_z^2+(2 n - 2 \text{sgn}(q) s_z + 1) |qB| + m^2\,,
\label{eq:energylevels-free}
\eeqn
where $n \geq 0$ is the Landau level, $s_z = -s, \dots, s$ is the projection of the spin $s$,
and $p_z$ is the particle's momentum along the magnetic field.
For vector meson $\rho$ with spin $s=1$,
its minimal effective mass square corresponding to the lowest energy state of
~\eq{eq:energylevels-free} with $p_z=0$, thus has the form of
\be
M_{\rho^\pm}^2(B) =  m_{\rho^\pm}^2 - e B \,.
\label{eq:mrhoB}
\ee
The vacuum masses of the $\rho^\pm$ mesons $m_\rho = 775.5\,\mbox{MeV}$
implies that the lowest energy of the charged $\rho$-meson in the external magnetic
field may become purely imaginary if the magnetic field exceeds the following critical
value
\be
e B_c = m_\rho^2 \simeq 0.6 {\rm GeV}^2 .
\label{eq:eBc}
\ee
This indicates that around this critical magnetic field, the vector mode will
become unstable, therefore there should appear electric charged vector meson
condensation in the vacuum, which is called the vacuum superconductor
\cite{Chernodub:2010qx}.

However, it is still not clear whether charged vector meson condensation
can happen, because different methods give different answers.

It was argued in \cite{Hidaka:2012mz} that due to Vafa-Witten theorem,
charged vector meson condensation cannot occur in QCD in a strong magnetic field
because any global-internal symmetry is not spontaneously broken by a magnetic field.
Then it was pointed out in \cite{Li:2013aa}, that the Vafa-Witten theorem would
not forbid the charged $\rho$ condensation under external magnetic fields.

However, the authors in \cite{Hidaka:2012mz} performed lattice calculation and
their results showed that the charged $\rho$ meson mass firstly decreases with
magnetic field and has a minimum around $eB\simeq 1 {\rm GeV}^2$, then again increases
with magnetic field, but the mass will not decrease to zero. This result is confirmed
in Ref.\cite{Andreichikov:2013zba} by solving the meson spectra in a relativistic
quark-antiquark system using the relativistic Hamiltonian technique. Also, in \cite{Wangkunlun:2013},
the author obtained similar results in the framework of Dyson-Schwinger equations.

On the other hand, the charged $\rho$ condensation was confirmed in SU(2) lattice
calculation in \cite{Braguta:2011hq}, and from calculations in the Nambu--Jona-Lasinio
(NJL) model \cite{Chernodub:2011mc,Frasca:2013kka} as well as from gauge/gravity
correspondence \cite{Callebaut:2011uc,Ammon:2011je}. However, the critical magnetic
field for charged $\rho$ mass becoming zero are different for different calculations.
The hadronic model gives $eB_c=0.6 {\rm GeV}^2$, the estimation by using NJL model in
\cite{Chernodub:2011mc} gives $eB_c> 1.0 {\rm GeV}^2$, while the NJL model calculation
in \cite{Frasca:2013kka} gives $eB_c \simeq 0.98 M_q^2$.

The motivation of our work is to investigate the vector meson carefully in the
NJL model and try to understand different results in different models. The paper
is organized as follows. In Sec. II, we give a general description of the NJL model
under magnetic field including the effective four-quark interaction in the vector
channel. In Sec. III, we introduce how to calculate the vector meson mass under
magnetic field. We give our numerical results and analysis in Sec. IV and then
in Sec. V we give the conclusion and discussion.

\section{The SU(2) NJL model under magnetic field}

We investigate the properties of vector meson under magnetic fields in the framework
of the SU(2) Nambu-Jona-Lasinio (NJL) model~\cite{Nambu:1961tp,
Nambu:1961fr, Klimt:1989pm, Vogl:1991qt, Klevansky:1992qe,Hatsuda:1994pi}.
The Lagrangian density of our model is given by
\begin{eqnarray}
{\cal{L}}&=&\bar{\psi}(i\not{\!\!D}-\hat{m})\psi+G_{S}\left[(\bar{\psi}\psi)^2
   +(\bar{\psi}i\gamma^5\vec{\tau}\psi)^2\right] \nonumber \\
 & & -G_{V}\left[(\bar{\psi}\gamma^{\mu}\mathbf{\tau}^a\psi)^2+
  (\bar{\psi}\gamma^{\mu}\gamma^5\mathbf{\tau}^a\psi)^2 \right] \nonumber \\
 & & -\frac{1}{4}F_{\mu\nu}F^{\mu\nu}.
\label{eq:L:basic}
\end{eqnarray}
Where $\psi$ corresponds to the quark field of two light flavors u and d,
$\hat{m}=\text{diag}(m_u,m_d)$ is the current quark mass matrix of u and d quarks, $\tau^a=(I,\vec{\tau})$ with $\vec{\tau}=(\tau^1,\tau^2,\tau^3)$ representing the
isospin Pauli
matrices, and $G_S$ and $G_V$ are the coupling constants with respect
to the scalar (pseudoscalar) and the vector (axial-vector) channels, respectively.
The covariant derivative, $D_{\mu}=\partial_{\mu}-i q_f A_{\mu}^{ext}$, couples quarks to an
external magnetic field $\bm{B}=(0,0,B)$ along the positive $z$ direction via a background
field, for example, $A_{\mu}^{ext}=(0,0,Bx,0)$. $q_f=(-1/3,2/3)$ is defined as the electric
charge of the quark field. The field strength tensor $F_{\mu\nu}$ is defined as usual by
$F_{\mu\nu}=\partial_{[\mu}A_{\nu]}^{ext}$, with $A_{\mu}^{ext}$ fixed as above.

The above Lagrangian is equivalent to the semi-bosonized Lagrangian
\begin{eqnarray}\label{NE2b}
{\cal{L}}_{sb}&=&\bar{\psi}(x)\left(i\gamma^{\mu}D_{\mu}-m_{0}\right)\psi(x)
-\bar{\psi}\left(\sigma+i\gamma_5\vec{\tau}\cdot\vec{\pi}\right)\psi\nonumber\\
&&-\frac{(\sigma^2+\vec{\pi}^2)}{4G_S}+\frac{(V_{\mu}^{a}V^{a\mu}+A_{\mu}^{a}A^{a\mu})}{4G_V}-\frac{B^2}{2},
\end{eqnarray}
where the Euler-Lagrange equations of motion for the auxiliary
fields lead to the constraints
\begin{eqnarray}\label{NE3b}
\sigma(x)&=&-2G_S\bar{\psi}(x)\psi(x),\\
\vec{\pi}(x)&=&-2G_S\bar{\psi}(x)i\gamma_5\vec{\tau}\psi(x), \\
V_\mu^a(x)&=&-2G_V\bar{\psi}(x)\gamma_\mu\tau^a \psi(x), \\
A_\mu^a(x)&=&-2G_V\bar{\psi}(x)\gamma_\mu\gamma^5\tau^a \psi(x).
\end{eqnarray}
In the vacuum, the quark-antiquark condenses and quarks obtain dynamical masses, so the
constituent quark mass $M$ of u,d quarks is defined by
\be \label{eq:gapequationmq}
M=m_0-2G_S <\bar{\psi}\psi>,
\ee
where we assume $m_u=m_d=m_0$. The inverse quark propagator takes the form of
\begin{eqnarray}\label{NE8b}
iS^{-1}_{Q}(\sigma,\vec{\pi})\equiv
i\gamma^{\mu}D_{\mu}-\left(\hat{m}+\sigma+i\gamma^{5}\vec{\tau}\cdot\vec{\pi}\right).
\end{eqnarray}
Under the external magnetic field $\mathbf{B}=(0,0,eB)$,
the solution of the Dirac equation with a constant magnetic field is known,
which forms the complete set of orthogonal wave-functions, and
the Ritus fermion propagator takes the form of
\cite{Ritus:1972ky,Fukushima:2009ft,Fayazbakhsh:2013cha}
\be\label{Ritus}
S_{Q}(x,y)=i\sum_{p=0}^{\infty}\hspace{-0.5cm}\int{\cal{D}}\tilde{p}~e^{-i\tilde{p}\cdot
(x-y)}P_{p}(x_{1})D_{Q}^{-1}(\bar{p})~P_{p}(y_{1}).
\ee
Where
\be
D_Q(\bar{p})=\gamma\cdot\bar{p}_Q-M,
\ee
with $M$ the constituent quark mass given in Eq.(\ref{eq:gapequationmq}), and
${\cal{D}}\tilde{p}\equiv
\frac{dp_{0}dp_{2}dp_{3}}{(2\pi)^{3}}$,
$\tilde{p}=(p_{0},0,p_{2},p_{3})$ and $\bar{p}$ is the Ritus
momentum
$$ \bar{p}=(p_{0},0,-s_{Q}\sqrt{2|QeB|p},p_{3}),$$
with $s_{Q}\equiv \mbox{sgn}(QeB)$, here $Q$ is the diagonal matrix, and we will use
$s\equiv \mbox{sgn}(q_{f}eB)$ for the elements of this $2\times 2$
matrix $s_{Q}$. The projection matrix with respect to the Dirac index according
to Ritus’ method is given below:
\begin{eqnarray}
\hspace{-0.8cm}P_{p}(x_{1})&\equiv&\frac{1}{2}[f_{p}^{+s}(x_{1})+\Pi_{p}f_{p}^{-s}(x_{1})]
\nonumber\\
&&\hspace{-0.2cm}+\frac{is_{Q}}{2}[f_{p}^{+s}(x_{1})-\Pi_{p}f_{p}^{-s}(x_{1})]
\gamma^{1}\gamma^{2}.
\end{eqnarray}
Here, $\Pi_{p}\equiv 1-\delta_{p,0}$ considers the spin degeneracy in the lowest Landau
level with $p=0$. Moreover, $f_{p}^{\pm s}(x_{1})$ are defined by
\begin{eqnarray}
\begin{array}{rclcrcl}
f_{p}^{+s}(x_{1})&\equiv&\phi_{p}\left(x_{1}-s_{Q}p_{2}\ell_{B}^{2}\right),&&
p&=&0,1,2,\cdots,\nonumber\\
f_{p}^{-s}(x_{1})&\equiv&\phi_{p-1}\left(x_{1}-s_{Q}p_{2}\ell_{B}^{2}\right),&&
p&=&1,2,3,\cdots,
\end{array}
\hspace{-0.4cm}\nonumber\\
\end{eqnarray}
where $\phi_{p}(x)$ is a function of Hermite polynomials $H_{p}(x)$
in the form
\begin{eqnarray}
\phi_{p}(x)\equiv a_{p}\exp\left(-\frac{x^{2}}{2\ell_{B}^{2}}\right)H_{p}\left(\frac{x}{\ell_{B}}\right).
\end{eqnarray}
Here, $a_{p}\equiv (2^{p}p!\sqrt{\pi}\ell_{B})^{-1/2}$ is the
normalization factor and $\ell_{B}\equiv |QeB|^{-1/2}$ is the
magnetic length.

\section{Vector meson mass under magnetic field}

\begin{figure}[ht]
\vspace*{-2truecm}
\centerline{\epsfxsize=15cm\epsffile{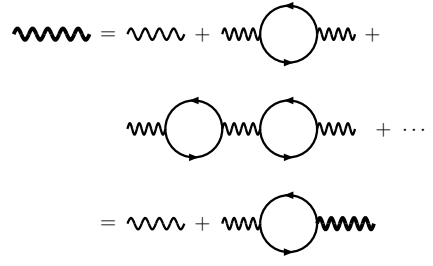}}
\vspace*{-14truecm}
\caption{The full propagator of $\rho$ meson in the
random phase approximation (RPA). Thick wavy lines indicate the full
propagator $D_{ab}^{\mu\nu}$ of $\rho$-meson, and thin
wavy lines the bare propagator $-2G_V\delta_{ab}$.}
\label{fig:RPArho}
\end{figure}

\begin{figure}[ht]
\vspace*{-6truecm}
\centerline{\epsfxsize=15cm\epsffile{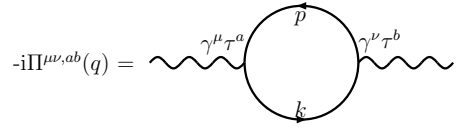}}
\vspace*{-13truecm}
\caption
{The $\rho$ meson polarization function $\Pi^{\mu\nu,ab}$ with one quark loop
contribution, i.e., the leading order contribution in $1/N_c$
expansion. }
\label{fig:Polarization}
\end{figure}

In the framework of the NJL model, mesons are $q \bar q $ bound states or resonances
and can be obtained from the quark-antiquark scattering amplitude
\cite{He:1997gn,Rehberg:1995nr}. In the random
phase approximation, the full propagator of $\rho$ meson shown in Fig.~\ref{fig:RPArho}
can be expressed to leading order in $1/N_c$ as an infinite sum of quark-loop chains,
and can also be recast into the form of a Schwinger-Dyson equation. The $\rho$-meson
propagator $D^{\mu\nu}_{ab}(q^2)$ can be obtained from one-loop polarization function
$\Pi_{\mu\nu,ab}(q^2)$ shown in Fig.~\ref{fig:Polarization} via the Schwinger-Dyson equation
and takes the form of
\begin{eqnarray}
\left[-iD_{ab}^{\mu\nu}\right]&=&\left[-2iG_V\delta_{ab}g^{\mu\nu}\right] + \nonumber \\
    & &  \left[-2iG_V\delta_{ac}g^{\mu\lambda}\right]
     \left[-i\Pi_{\lambda\sigma,cd}\right]\left[-iD^{\sigma\nu}_{db}\right],
\end{eqnarray}
where $a,b,c,d$ are isospin indices, and $\mu$, $\nu$ Lorentz indices.

The one quark loop polarization function $\Pi_{\mu\nu,ab}(q^2)$ takes the form of
\begin{eqnarray}
\Pi^{\mu\nu,ab}(q^2) & = & i\sum_{p,k=0}^{\infty}\int\ {\cal{D}}\tilde{p}\ {\cal{D}}\tilde{k}
                 \int d^{4}x \nonumber\\
 & & e^{-i(\tilde{p}-\tilde{k}-q)\cdot x}\Lambda^{\mu\nu,ab}_{pk}(\bar{p},\bar{k},x_1),
\end{eqnarray}
where
\begin{eqnarray}
\Lambda^{\mu\nu,ab}_{pk}(\bar{p},\bar{k},x_1)&=&
\mbox{tr}_{sfc}\big[\gamma^{\mu}\mathbf{\tau}^{a}P_{p}(x_{1})D^{-1}_{Q}(\bar{p})P_{p}(0)
\gamma^{\nu}\mathbf{\tau}^{b}
\nonumber\\
&&\times K_{k}(0)D^{-1}_{Q}(\bar{k})K_{k}(x_{1})\big]
\end{eqnarray}
with external momentum $q=p-k$.

\subsection{Charged $\rho^{\pm}$ meson}

For charged $\rho^{\pm}$ meson, the isospin Pauli matrices take $\mathbf{\tau}^{a}=\mathbf{\tau}^{\pm}$ and $\mathbf{\tau}^{b}=\mathbf{\tau}^{\mp}$,
respectively, with
\begin{eqnarray}
\mathbf{\tau}^{\pm}=\frac{1}{\sqrt{2}}(\mathbf{\tau}^{1}\pm \mathbf{\tau}^{2}).
\end{eqnarray}
In the rest frame, $q_{\mu}=(M_{\rho^{\pm}},\mathbf{0})$, the polarization function
has the form of
\begin{eqnarray}
\Pi^{\mu\nu}_{\rho^{\pm}}(q^2)&=&i\sum_{p,k=0}^{\infty}\int \frac{dk_{0}dk_{3}}{(2\pi)^3}\int dk_2 dx_1 \Lambda^{\mu\nu}_{\rho^{\pm},{pk}}(\bar{p},\bar{k},x_1),\nonumber \\
\end{eqnarray}
where
\begin{eqnarray}
\Lambda^{\mu\nu}_{\rho^{\pm},{pk}}(\bar{p},\bar{k},x_1)&=& Tr_{sfc}[\gamma^{\mu}\mathbf{\tau}^{\pm}P_p(x_1)D_q^{-1}(\bar{p})P_p(0) \nonumber \\
& & \gamma^{\nu}\mathbf{\tau}^{\mp}K_k(0)D_Q^{-1}(\bar{k})K_k(x_1)].
\end{eqnarray}

In the following, we calculate the components of polarization tensor for $\rho^{\pm}$.
$\mu,\nu=1,2$ correspond to the transverse components of the polarization tensor and
we introduce the definition $\mu_{\bot}\nu_{\bot}$. Thus,
\begin{eqnarray}
\Lambda^{\mu_{\bot}\nu_{\bot}}_{\rho^{\pm},{pk}}& = & Tr_{scf}[(A^{+}-is_p\gamma^1\gamma^2A^{-})\gamma^{\mu}
      D_Q^{-1}(\bar{p})\gamma^{\nu} \nonumber \\
& & (\alpha^+-is_p\gamma^1\gamma^2\alpha^-)D_Q^{-1}(\bar{k})],
\end{eqnarray}
where
\beqn
&&A^{\pm}=\frac{1}{2}(f_p^{+s_p}(x_1)f_k^{+s_k}(x_1)\pm\Pi_{p}\Pi_{k}f_p^{-s_p}(x_1)f_k^{-s_k}(x_1))\nonumber\\
&&\alpha^{\pm}=\frac{1}{2}(f_p^{+s_p}(0)f_k^{+s_k}(0)\pm\Pi_{p}\Pi_{k}f_p^{-s_p}(0)f_k^{-s_k}(0))\nonumber
\eeqn
and $s_p=\text{sgn}(q_{fp} eB)$( $ q_{fp}$ is the electric charge of u quark for $\rho^{+}$, and
$ q_{fp}$ is the electric charge of d quark for $\rho^{-}$ ).

The components can be derived and have the following forms:
\begin{eqnarray}
\Lambda^{11}_{\rho^{\pm},{pk}}&=& 4N_cN_f\frac{1}{(p_0^2-\omega_p^2)
         (k_0^2-\omega_k^2)} [(\bar{p}\bar{k}-M^2)A^+\alpha^+ \nonumber \\
       & & +(2\bar{k}_2\bar{p}_2
       +\bar{p}\bar{k}-M^2)A^-\alpha^-] ,\\
\Lambda^{12}_{\rho^{\pm},{pk}}&=& 4N_cN_f\frac{1}{(p_0^2-\omega_p^2)(k_0^2-\omega_k^2)} \nonumber \\
    & & [(\bar{p}\bar{k}+2\bar{p}_2\bar{k}_2-M^2)(-is_pA^-\alpha^+) \nonumber\\
    & & +(\bar{p}\bar{k}-M^2)(-is_pA^+\alpha^-)], \\
\Lambda^{21}_{\rho^{\pm},{pk}}&=& 4N_cN_f\frac{1}{(p_0^2-\omega_p^2)(k_0^2-\omega_k^2)}\nonumber \\
  & & [(-\bar{p}\bar{k}+M^2)(-is_pA^-\alpha^+)+(-\bar{p}\bar{k} \nonumber\\
  & &  -2\bar{p}_2\bar{k}_2+M^2)(-is_pA^+\alpha^-)], \\
\Lambda^{22}_{\rho^{\pm},{pk}}&=& 4N_cN_f\frac{1}{(p_0^2-\omega_p^2)
   (k_0^2-\omega_k^2)} [(\bar{p}\bar{k}-M^2)A^-\alpha^- \nonumber\\
& & +(2\bar{p}_2\bar{k}_2+\bar{p}\bar{k}-M^2)A^+\alpha^+],
\end{eqnarray}
where the $\omega_p^2=2|q_{fp} eB|p+k_3^2+M^2$.

For $\mu,\nu=0,3$, they correspond to the longitudinal components of the polarization tensor and are defined by $\mu_{\|}\nu_{\|}$.
\begin{eqnarray}
\Lambda^{\mu_{\|}\nu_{\|}}_{\rho^{\pm},{pk}}
   &=& Tr_{scf}[(B^{+}-is_p\gamma^1\gamma^2B^{-})\gamma^{\mu}D_Q^{-1}(\bar{p}) \nonumber \\
   & & \gamma^{\nu}({\beta}^{+}-is_p\gamma^1\gamma^2{\beta}^-)D_Q^{-1}(\bar{k})],
\end{eqnarray}
where
\beqn
&&B^{\pm}=\frac{1}{2}[\Pi_{p}f_k^{s_k}(x_1)f_p^{-s_p}(x_1)\pm\Pi_{k}f_p^{s_p}(x_1)f_k^{-s_k}(x_1)]\nonumber\\
&&\beta^{\pm}=\frac{1}{2}[\Pi_{p}f_k^{s_k}(0)f_p^{-s_p}(0)\pm\Pi_{k}f_p^{s_p}(0)f_k^{-s_k}(0)].\nonumber
\eeqn
The component
\begin{eqnarray}
\Lambda^{33}_{\rho^{\pm},{pk}}& =& 4N_cN_f\frac{1}{(p_0^2-\omega_p^2)(k_0^2-\omega_k^2)} \nonumber \\
& & [(p_0k_0-\bar{p}_2\bar{k}_2+k_3^2-M^2)B^+\beta^+ \nonumber \\
& & + (p_0k_0+\bar{p}_2\bar{k}_2+k_3^2-M^2)B^-\beta^-].
\end{eqnarray}
The other matrix elements of $\Pi^{\mu\nu}_{\rho^{\pm}}$ are zero.

Finally, we can get the matrix
\beqn \label{eq:matrixrhopm}
\Pi^{\mu\nu}_{\rho^{\pm}}=\left(\begin{matrix}0&0&0&0\cr0&\Pi^{11}&\Pi^{12}&0 \cr0& \Pi^{21}&\Pi^{22}&0\cr0&0&0&\Pi^{33} \end{matrix}\right)
=\left(\begin{matrix}0&0&0&0\cr0&a& -ib &0 \cr0& ib&a&0\cr0&0&0&c \end{matrix}\right),
\eeqn
where we have used relations $\Pi^{11}=\Pi^{22}=a$ and $\Pi^{12}=-\Pi^{21}=ib$.

The polarization tensor can be decomposed as following
\be
\Pi^{\mu\nu}_{\rho^{\pm}}=[A_1^2 P_{1}^{\mu\nu}+A_2^2 P_{2}^{\mu\nu}+A_3^2 L^{\mu\nu}+A_4^2u^{\mu}u^{\nu}],
\ee
where $u^{\mu}=(1,0,0,0)$ is the four momentum in the rest frame, and
we have introduced the spin projection operator
\beqn
P_{1}^{\mu\nu} & = & -\epsilon_1^{\mu}\epsilon_1^{\nu}, \, (s_z= -1\, \text{for} \, \rho^\pm), \\
P_{2}^{\mu\nu} & = & -\ep_2^{\mu}\ep_2^{\nu}, \, (s_z= 1\, \text{for} \, \rho^\pm),\\
L^{\mu\nu} & = & -b^{\mu}b^{\nu}, \, (s_z=0 \, \text{for} \, \rho^\pm).
\eeqn
Here the right and left-handed polarization vectors
\beqn
\ep_1^{\mu} &=& \frac{1}{\sqrt{2}}(0,1,i,0), \\
\ep_2^{\mu} &=& \frac{1}{\sqrt{2}}(0,1,-i,0),
\eeqn
are parallel or anti-parallel to the the external magnetic field direction
$b^{\mu}=(0,0,0,1)$.

Consequently, the charged $\rho^{\pm}$ meson propagator can be written as:
\beqn
D_{\rho^\pm}^{\mu\nu}(q^2) &=&[D_1(q^2)P_{1}^{\mu\nu}+D_2(q^2)P_{2}^{\mu\nu} \nonumber \\
            & & +D_3(q^2)L^{\mu\nu}+D_4(q^2)u^{\mu}u^{\nu}].
\eeqn
Each component $D_i$ can be written in the form of
\be
D_i(q^2)=\frac{2G_V}{1+2G_V A_{i}^2}
\ee
by using RPA, and the mass of charged $\rho^{\pm}$ can be determined by the gap equation:
\be
1+2G_V A_{i}^2=0.
\ee

For charged $\rho^\pm$ meson with spin components $s_z=1,0,-1$, we have gap equations:
\beqn \label{eq:gapequationrho}
& & 1+2G_V A_{1}^2=0, \, (s_z=-1), \\
& & 1+2G_V A_{2}^2=0, \, (s_z=1), \\
& & 1+2G_V A_{3}^2=0, \, (s_z=0).
\eeqn
From Eq.(\ref{eq:matrixrhopm}), it is easy to find that
\beqn\label{A^2}
&&A^2_{1}=-(a+b),  \\
&&A^2_{2}=b-a, \nonumber\\
&&A^2_{3}=c.
\eeqn
In the rest frame of $\rho$ with $q^{\mu}=(M_{\rho^{\pm}},\mathbf{0})$,
it is easy to find $A_{4}^2=0$, which is required by the Ward identity.

\subsection{Neutral $\rho^0$ meson}

For charge neutral $\rho^0$ meson, the isospin Pauli matrices take $\mathbf{\tau}^{a}=\mathbf{\tau}^3$ and $\mathbf{\tau}^{b}=\mathbf{\tau}^3$.
In the rest frame, $q_{\mu}=(M_{\rho^{0}},\mathbf{0})$, the polarization function
has the form of
\begin{eqnarray}
& & \Pi^{\mu\nu}_{\rho^{0}}(q^2)= i\sum_{p,k=0}^{\infty}\int \frac{dk_{0}dk_{3}}{(2\pi)^3}
T^{\mu\nu}_{\rho^{0}}(\bar{p},\bar{k}), \nonumber \\
& & T^{\mu\nu}_{\rho^{0}}(\bar{p},\bar{k}) = \int dk_2dx_1\Lambda^{\mu\nu}_{\rho^{0},{pk}},
\end{eqnarray}
with
\begin{eqnarray}
\Lambda^{\mu\nu}_{\rho^{0},{pk}}(\bar{p},\bar{k},x_1)&=& Tr_{sfc}[\gamma^{\mu}\mathbf{\tau}^{3}P_p(x_1)D_q^{-1}(\bar{p})P_p(0) \nonumber \\
& & \gamma^{\nu}\mathbf{\tau}^{3}K_k(0)D_Q^{-1}(\bar{k})K_k(x_1)].
\end{eqnarray}

By using the orthonormality relations of $f_{p}^{\pm s}$:
\begin{eqnarray}
\int dx_{1}\ f^{+s}_{p}(x)f^{+s}_{k}(x)\bigg|_{p_2=k_2}&=&\delta_{pk},\nonumber\\
\end{eqnarray}
and
\begin{eqnarray}
\int dk_2\ f^{+s}_{p}(0)f^{+s}_{k}(0)\bigg|_{p_2=k_2}&=&
\frac{\delta_{pk}}{\ell_B^2},\nonumber\\
\end{eqnarray}
$T^{\mu\nu}_{\rho^{0}}$ can be simplified as
\beqn
& &T^{11}_{\rho^{0}}(\bar{p},\bar{k})=T^{22}_{\rho^{0}}(\bar{p},\bar{k}) =
4N_c\sum_{q_f\in\{\frac{-1}{3},\frac{2}{3}\}}|q_f eB| \nonumber\\
& &
  \times \frac{p_0k_0-k_3^2-M^2}{(p_0^2-\omega_p^2)(k_0^2-\omega_k^2)}
  (\frac{1}{2}\delta_{k,p-1}+\frac{1}{2}\delta_{p,k-1}),\\
& & T^{33}_{\rho^{0}}(\bar{p},\bar{k})=2N_c\sum_{q_f\in\{\frac{-1}{3},\frac{2}{3}\}}|q_f eB| \nonumber \\ & & \times \frac{p_0k_0-\bar{p}_2\bar{k}_2+k_3^2-M^2}{(p_0^2-\omega_p^2)
(k_0^2-\omega_k^2)}{\alpha}_k{\delta}_{p,k},
\eeqn
where $\alpha_k=2-\delta_{k,0}$.

We get the matrix
\beqn \label{eq:matrixrho0}
\Pi^{\mu\nu}_{\rho^{0}}=\left(\begin{matrix}0&0&0&0\cr0&\Pi^{11}_{\rho^0}&0&0 \cr0& 0&\Pi^{22}_{\rho^0}&0\cr0&0&0&\Pi^{33}_{\rho^0} \end{matrix}\right)
=\left(\begin{matrix}0&0&0&0\cr0&d& 0 &0 \cr0& 0&d&0\cr0&0&0&e \end{matrix}\right),
\eeqn
here we have used the relations $\Pi^{11}_{\rho^0}=\Pi^{22}_{\rho^0}=d$ and $\Pi^{33}_{\rho^0}=e$.

Similar to the case of charged $\rho^\pm$ mesons, charge neutral $\rho^0$ meson with spin
components $s_z=1,0,-1$ take the following gap equations:
\beqn
& & 1+2G_V A_{1}^2=0, \, (s_z=-1), \\
& & 1+2G_V A_{2}^2=0, \, (s_z=1), \\
& & 1+2G_V A_{3}^2=0, \, (s_z=0),
\eeqn
From Eq.(\ref{eq:matrixrho0}), it is easy to find that
\beqn\label{A^2}
&&A^2_{1}=-d,  \\
&&A^2_{2}=-d, \nonumber\\
&&A^2_{3}=e.
\eeqn

\section{Numerical results}

Following Ref.\cite{He:1997gn}, the model parameters are fitted by reproducing the pion
decay constant, the vacuum quark mass,
the mass of $\pi$ and the mass of $\rho$ in the vacuum. They are given by $\Lambda = 582 {\rm MeV}, G_S{\Lambda}^2 = 2.388,G_V{\Lambda}^2=1.73$. These parameters correspond to $f_{\pi} = 95 {\rm MeV}$, $m_{\pi}=140 {\rm MeV}, M_{\rho}=768 {\rm MeV} $, the vacuum condensation $<\bar{u}u>=-(252)^3 {\rm MeV}^3$, the vacuum quark mass $M=458 {\rm MeV}$ and the current quark mass $m_0= 5 {\rm MeV}$.
We use the soft cut-off function as in \cite{Frasca:2011zn}
\beqn
&&f_\Lambda=\frac{\Lambda^{10}}{\Lambda^{10}+\mathbf{k}^{2*5}}, \\
&&f_{\Lambda,eB}=\frac{\Lambda^{10}}{\Lambda^{10}+(k_z^2+2|Q_f eB|k)^5},
\eeqn
for zero and nonzero magnetic fields.

\subsection{Masses of charged and neutral $\rho$ mesons}

\begin{figure}[!thb]
\centerline{\includegraphics[width=8cm]{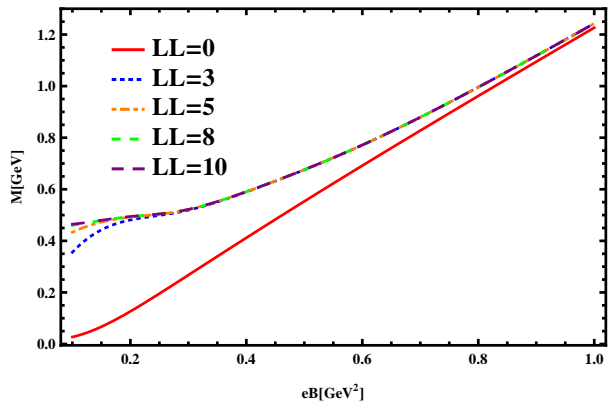}}
\caption{Quark constitute mass $M$ as a function of $eB$ with different Landau levels included in the numerical caculations.}
\label{fig:quarkmass}
\end{figure}

We calculate the vector meson mass numerically, and the quark mass $M$ is also solved
self-consistently from the gap equation Eq. (\ref{eq:gapequationmq}). It can be
read from Fig. \ref{fig:quarkmass}, the constituent quark mass increases with magnetic
field, which is the well-known magnetic catalysis effect \cite{Gusynin:1995nb}.

\begin{figure}[!thb]
\centerline{\includegraphics[width=8cm]{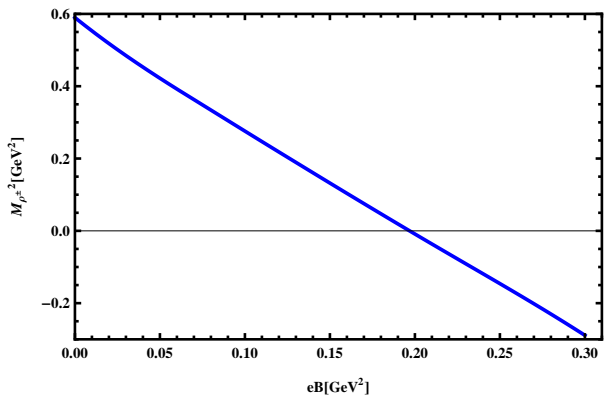}}
\caption{The mass square of charged $\rho^\pm$ with spin component $s_z=\pm1$ as a function
of $eB$.}
\label{fig:chargedrhomesonsz}
\end{figure}

The numerical results for the mass square of charged $\rho^\pm$ with spin component $s_z=\pm1$
is shown in Fig.~\ref{fig:chargedrhomesonsz}. For numerical calculations, we have summed 20
Landau levels, the results for summation above 10 Landau levels are saturated. It is found
that even though the constituent quark mass increases with magnetic field, the mass square of
charged $\rho^\pm$ with $s_z=\pm1$ decreases with $eB$ linearly
$M_{\rho^\pm}^2(B) =  m_{\rho^\pm}^2 - \kappa e B$ with the slope $\kappa \approx 3$,
and the mass square goes to zero at $eB_c \simeq 0.2 {\rm GeV}^2$.
In Appendix \ref{appendix:kappa}, by performing the weak magnetic expansion we obtain
$M_{\rho^\pm}^2(B) =  m_{\rho^\pm}^2- \kappa e B$ with $\kappa_w \approx 1.92$. For both
cases, the mass square of charged $\rho^\pm$ with spin component $s_z=\pm1$ decreases
faster with $eB$ than the case of free charged relativistic particle, where
$M_{\rho^\pm}^2(B) =  m_{\rho^\pm}^2- \kappa e B$ with $\kappa=1$ which gives
the critical magnetic field $eB_c \approx 0.6 {\rm GeV}^2$.

\begin{figure}
\centerline{\includegraphics[width=8cm]{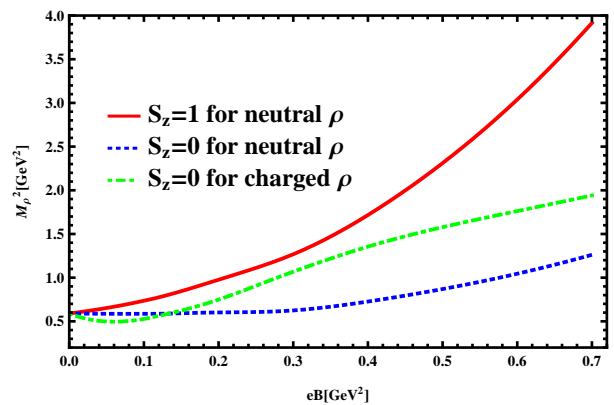}}
\caption{Masses of charged vector meson $\rho^\pm$ with $s_z=0$ and
neutral vector meson $\rho_0$ with $s_z=0,\pm1$ as functions of magnetic field $eB$.}
\label{fig:neutralrho}
\end{figure}

Fig.~\ref{fig:neutralrho} shows the numerical results for masses of charged vector meson
$\rho^\pm$ with $s_z=0$ and neutral vector meson $\rho_0$ with $s_z=0,\pm1$ as functions
of magnetic field $eB$. It is found that for all these modes, the masses of vector mesons
increase with magnetic field. The neutral vector meson $\rho^0$ with $s_z=\pm1$ increase fast
with the magnetic field. The neutral vector meson $\rho^0$ with $s_z=0$ almost
remains as a constant at low magnetic field, but then slowly increases with the magnetic field
when $eB>0.3 {\rm GeV}^2$. The lattice group \cite{Luschevskaya:2014tza} found that
the neutral vector meson $\rho^0$ with $s_z=\pm1$ also increase
with the magnetic field, however, their result on neutral vector meson $\rho^0$ with
$s_z=0$ decreases with the magnetic field.

\subsection{Mass of charged $\rho^{\pm}$ with $s_z=\pm1$ at LLL}

As mentioned in the Introduction that there some other results showed that
the mass of the charged vector meson $\rho^\pm$ can never drop to zero at high
magnetic field. For example, \cite{Hidaka:2012mz} performed lattice calculation and
their results showed that the charged $\rho$ meson mass firstly decreases with
magnetic field and has a minimum around $eB\simeq 1 {\rm GeV}^2$, then again increases
with magnetic field, and this result was confirmed in Ref.\cite{Andreichikov:2013zba}
by solving the meson spectra in a relativistic quark-antiquark system using the relativistic
Hamiltonian technique. Also, in \cite{Wangkunlun:2013},
the author obtained similar result in the framework of Dyson-Schwinger equations.

\begin{figure}[!thb]
\centerline{\includegraphics[width=8cm]{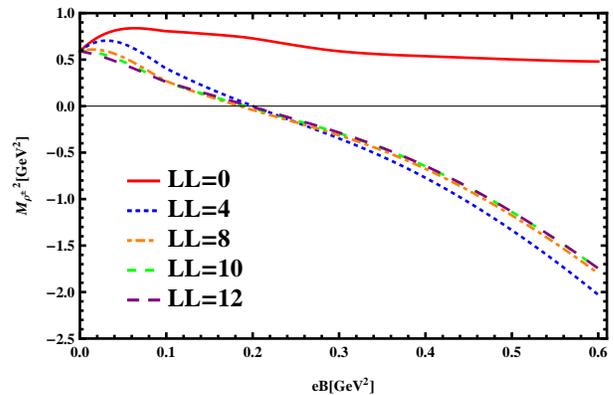}}
\caption{$M_{\rho^\pm}^2$ for $s_z=1$ as a function of $eB$ with different Landau levels,
with LL the Landau Levels.}
\label{fig:mrhoLLL}
\end{figure}

In this section, we analyze possible reasons. We don't know the detailed
calculations from lattice in Ref.\cite{Hidaka:2012mz} and DSE Ref.\cite{Wangkunlun:2013}.
However, the authors used lowest Landau level approximation in Ref.\cite{Andreichikov:2013zba}.
As we all know, when the magnetic field becomes strong, there exits dimensional reduction
$D \rightarrow D-2$, and in strong magnetic field limit, lowest Landau level (LLL)
approximation is widely used. From our numerical results in Fig.\ref{fig:chargedrhomesonsz},
the charged $\rho^\pm$ for $s_z=1$ drops to zero at $eB_c=0.2 {\rm GeV}^2$, which is not
in the range of strong magnetic field. Therefore, we investigate how LLL approximation
will affect the mass of $\rho$ meson.

Fig.~\ref{fig:quarkmass} shows the constitute quark mass with different Landau levels,
where $LL=0$ indicates the LLL approximation result. It is observed that at weak magnetic
field, the LLL approximation gives a rather small constituent quark mass and cannot describe
the spontaneous chiral symmetry breaking. At high magnetic fields, the LLL approximation
gives almost the same constituent quark mass as summing over higher Landau levels.

The mass square of charged $\rho^\pm$ for $s_z=\pm1$ with different Landau levels is given in
Fig. \ref{fig:mrhoLLL}. Contrary to the case of constituent quark mass, at weak magnetic
field, LLL approximation gives similar results of $M_{\rho^\pm}^2 (s_z=\pm1)$ comparing with
higher Landau level summation results. However, when the magnetic field increases, the
difference between $M_{\rho^\pm}^2 (s_z=\pm1)$ at LLL approximation and higher LL summation
becomes larger and larger. It is noticed that $M_{\rho^\pm}^2 (s_z=\pm1)$ at LLL approximation
changes flatly with the increase of magnetic field and does not drop to zero! This result
qualitatively agrees with the results in
Ref.\cite{Hidaka:2012mz,Andreichikov:2013zba,Wangkunlun:2013}.

\begin{figure}[!thb]
\centerline{\includegraphics[width=8cm]{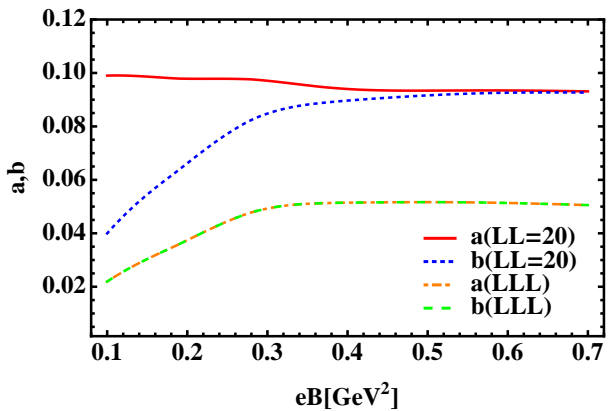}}
\caption{$a,b$ as functions of $eB$ with Landau levels LL=0 and 20, respectively. We
have taken $M_{\rho^+}=768 {\rm MeV}$.}
\label{fig:abeB}
\end{figure}

Let's further analyze why $M_{\rho^\pm}^2 (s_z=\pm1)$ does not drop to zero by using
LLL approximation in our framework. For $\rho^+(s_z=1)$, the mass is solved from the
gap equation $1-2G_V(a+b)=0$. We plot $a$ and $b$ for $M_{\rho^+}=768 {\rm MeV}$
in Fig. \ref{fig:abeB} as functions of $eB$ with Landau levels LL=0 and 20, respectively.
It is found that at weak magnetic field, the LLL approximation only gives $1/5$
contribution to $a$ and $1/2$ contribution to $b$. At high magnetic field, the LLL
approximation gives almost $1/2$ contribution to both $a$ and $b$.

\begin{figure}[!thb]
\centerline{\includegraphics[width=8cm]{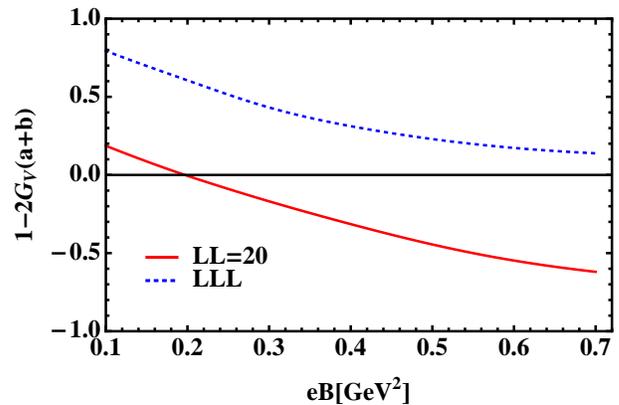}}
\caption{The gap function $1-2G_V(a+b)$ at $M_{\rho^+}=0$ as a function of $eB$
with Landau levels LL=0 and 20, respectively.  }
\label{fig:gapfunctioneB}
\end{figure}

We set $M_{\rho^+}=0$, and in Fig. \ref{fig:gapfunctioneB}
we plot the gap function $1-2G_V(a+b)$ as a function of $eB$ with Landau levels
LL=0 and 20, respectively. We can see that the gap function at LLL approximation
does not cross zero axis, however, the gap function at LL=20 crosses zero axis
at $eB_c\simeq 0.2 {\rm GeV}^2$, which is exactly the results given by solving
the pole mass from the gap equation.

\section{Conclusion}

After the prediction of the vacuum superconductor based on the
the energy of a free particle under magnetic fields by neglecting the internal
structure of vector mesons \cite{Chernodub:2010qx,Chernodub:2011mc},
there are more efforts trying to investigate properties of charged $\rho$ meson
by considering the internal structure of vector mesons. However, most of these
calculations tend to conclude that the mass of the charged vector meson $\rho^\pm$
will not drop to zero at high magnetic field, therefore there would be no vacuum
superconductor \cite{Hidaka:2012mz,Andreichikov:2013zba,Wangkunlun:2013}. Therefore
in this work, we carefully investigated the charged and neutral $\rho$ meson mass
in the presence of external magnetic field in the framework of NJL model.

In the NJL model, mesons are constructed by infinite sum of quark-loop chains by
using random phase approximation. We calculate the $\rho$ meson polarization tensor
to the leading order of $1/N_c$ expansion by considering one quark loop contribution,
and solve the masses of vector meson with different spin component from gap equations.
The constituent quark mass is also solved self-consistently under the magnetic field.
It is found that the constituent quark mass increases with the magnetic field, which
is famous magnetic catalysis effect. The masses of the neutral vector meson $\rho^{0}$
with spin component $s_z=0,\,\pm1$ and the charged vector meson $\rho^{\pm}$ with $s_z=0$
also increases with magnetic field. However, the mass square of the charged vector meson
$\rho^{+}$ ($\rho^{-}$) with $s_z=+1$ ($s_z=-1$) decreases linearly with magnetic field
and drops to zero at the critical magnetic field $e B_c \simeq 0.2 {\rm GeV}^2$, which
indicates the possible condensation of charged vector meson in the vacuum. This critical
magnetic field is much lower than the value $eB_c=0.6 {\rm GeV}^2$ predicted by a
point-like vector meson.

At the end, we analyze possible reasons why other groups \cite{Hidaka:2012mz,Andreichikov:2013zba,Wangkunlun:2013} obtained different results
on the charged $\rho$ meson mass. One possible reason is that it might due to the lowest
Landau level approximation as used in \cite{Andreichikov:2013zba} (though we are not sure
whether LLL approximation is used in \cite{Hidaka:2012mz,Wangkunlun:2013}). We find
that if we use lowest Landau level approximation, the mass of the
charged vector meson $\rho^{\pm}$ for $s_z=\pm1$ cannot drop to zero at high magnetic fields!
Another reason might due to the spin decomposition of the $\rho$ meson polarization tensor.
In order to obtain conclusive results on the masses of charged vector meson under magnetic
field, more efforts are needed in the future.

\vskip 1 cm
{\bf Acknowledgement.---} We thank J.Y.Chao, M. Chernodub, J. V. Doorsselaere,
M. Frasca and D.N.Li for valuable discussions. This work is supported by the NSFC under
Grant No. 11275213, DFG and NSFC (CRC 110), CAS key project KJCX2-EW-N01, and Youth
Innovation Promotion Association of CAS. L.Yu is partially supported by China
Postdoctoral Science Foundation under Grant No. 2014M550841.

\appendix

\section{Integrals}

For numerical calculations, we have defined the notations for the integrals of $k_0$ as follows:
\beqn
\label{intk_0}
& & i\int\frac{dk_0}{2\pi}\frac{\bar{p}\bar{k}-M^2}{(p_0^2-\omega_p^2)(k_0^2-\omega_k^2)}
 =\frac{1}{2}(iI_1+iI_1^{'}) \nonumber \\
 && \,~~~~ -\left[\frac{1}{2}\left(M^2-{\bar{p}_2}^2
-{\bar{k}_2}^2\right)+\bar{p}_2\bar{k}_2\right]iI_2, \\
& & i\int\frac{dk_0}{2\pi}\frac{\bar{p}\bar{k}-M^2+2\bar{p}_2\bar{k}_2}
{(p_0^2-\omega_p^2)(k_0^2-\omega_k^2)}
 =\frac{1}{2}(iI_1+iI_1^{'}) \nonumber \\
& & \,~~~~ -\left[\frac{1}{2}\left(M^2-{\bar{p}_2}^2
-{\bar{k}_2}^2\right)-\bar{p}_2\bar{k}_2\right]iI_2,  \\
&& i\int\frac{dk_0}{2\pi}\frac{p_0k_0+\bar{p}_2\bar{k}_2+k_3^2-M^2}{(p_0^2-\omega_p^2)(k_0^2-\omega_k^2)}
=\frac{1}{2}(iI_1+iI_1^{'}) \nonumber \\
& & \,~~~~ +\left[2k_3^2-\frac{1}{2}\left(M^2-{\bar{p}_2}^2-{\bar{k}_2}^2\right)+\bar{p}_2\bar{k}_2\right]iI_2,
\\
&&i\int\frac{dk_0}{2\pi}\frac{p_0k_0-\bar{p}_2\bar{k}_2+k_3^2-M^2}{(p_0^2-\omega_p^2)(k_0^2-\omega_k^2)} =\frac{1}{2}(iI_1+iI_1^{'}) \nonumber \\
& & \,~~~~ +\left[2k_3^2-\frac{1}{2}\left(M^2-{\bar{p}_2}^2-{\bar{k}_2}^2\right)
-\bar{p}_2\bar{k}_2\right]iI_2, \\
\eeqn
with
\beqn
&&I_1=\int\frac{dk_0}{2\pi}\frac{1}{k_0^2-\omega_p^2}, \\
&&I_1^{'}=\int\frac{dk_0}{2\pi}\frac{1}{k_0^2-\omega_k^2}, \\
&&I_2=\int\frac{dk_0}{2\pi}\frac{1}{((k_0+q_0)^2-\omega_p^2)(k_0^2-\omega_k^2)}.
\eeqn

Replacing the integral over $k_0$ to Matsubara summation as in \cite{Rehberg:1995nr}, one obtains:
\beqn
&&iI_1=\frac{\tanh[\frac{\omega_p}{2T}]}{2\omega_p}, \\
&&iI_1^{'}=\frac{\tanh[\frac{\omega_k}{2T}]}{2\omega_k}, \\
&&iI_2=-\int dy\frac{\tanh[\frac{\sqrt{(-1+y)(M^2y-\omega_k^2)+y\omega_p^2}}{2T}]}{4[(-1+y)
(M^2y-\omega_k^2)+M^2\omega_p^2]^\frac{3}{2}}.
\eeqn
At zero temperature $T=0$, we have
\beqn
&&\tanh[\frac{\omega_p}{2T}]=1, \\
&&\tanh[\frac{\omega_k}{2T}]=1, \\
&&\tanh[\frac{\sqrt{(-1+y)(M^2y-\omega_k^2)+y\omega_p^2}}{2T}]=1.
\eeqn

\section{The calculation of $\Pi^{\mu\nu}_{ab}$ in weak magnetic field}
\label{appendix:kappa}

In this Appendix, we investigate whether the mass of charged $\rho$ meson
decreases with $eB$ at weak magnetic field. In order to double check our results
by using the Ritus propagator, here we use the quark
propagator with the form in Ref. \cite{Gusynin:1995nb},
\be
\label{propagator}
\tilde{S}(k)=i \exp(-\frac{\textbf{k}_{\bot}^2}{|QeB|})\sum_{n=0}^{\infty}(-1)^n
\frac{D_n(QeB,k)}{k_0^2-2|QeB|n-k_3^2-M^2}
\ee
with
\beqn
D_n(QeB,k)&=&(k^0\gamma^0-k^3\gamma^3+m) \nonumber\\
&& [(1-i\gamma^1\gamma^2sgn(QeB))L_n(2\frac{\textbf{k}_\bot^2}{|QeB|})\nonumber\\
&&-(1+i\gamma^1\gamma^2sgn(QeB))L_{n-1}(2\frac{\textbf{k}_\bot^2}{|QeB|})]\nonumber\\
&&+4(k^1\gamma^1+k^2\gamma^2)L_{n-1}^1(2\frac{\textbf{k}_\bot^2}{|QeB|}).
\eeqn
where $\textbf{k}_\bot$ is ($k^1$, $k^2$), $L_n$ is Laguerre polynomials and
Q is a diagonal matrix with the entries $q_f=\{2/3$,$-1/3\}$.

The one-loop polarization of $\rho$ meson is $\Pi^{\mu\nu,ab}=-iTr[\tilde{S}(k)\gamma^{\mu}\tau_a\tilde{S}(p)\gamma^{\nu}\tau_b]$,
where $p=k+q$ and $q$ is the momentum of $\rho$. We only calculate the polarization function
of $\rho^-$ in the rest frame,
\beqn
& & \Pi^{11}=iN_cN_f\int\frac{dk^4}{(2\pi)^4}\exp\left(-\frac{9\textbf{k}_\bot^2}{2|eB|}\right)
\sum_{k=0}^{\infty}\sum_{p=0}^{\infty}(-1)^{p+k} \nonumber\\&&\left[8\left(L_k(3\frac{\textbf{k}_\bot^2}{|eB|})L_p
(6\frac{\textbf{k}_\bot^2}{|eB|})+L_{k-1}(3\frac{\textbf{k}_\bot^2}
{|eB|})L_{p-1}(6\frac{\textbf{k}_\bot^2}{|eB|})\right)
\right] \nonumber\\
&&\frac{k_0p_0-k_3^2-M^2}{(k_0^2-\frac{4}{3}|eB|k-k_3^2-M^2)(p_0^2-\frac{2}{3}|eB|p-p_3^2-M^2)},
\eeqn
\beqn
& & \Pi^{12}=iN_cN_f\int\frac{dk^4}{(2\pi)^4}\exp\left(-\frac{9\textbf{k}_\bot^2}{2|eB|}\right)\sum_{k=0}^{\infty}\sum_{p=0}^{\infty}(-1)^{p+k} \nonumber\\&&\left[8i\left(L_{k-1}(3\frac{\textbf{k}_\bot^2}{|eB|})L_{p-1}
(6\frac{\textbf{k}_\bot^2}{|eB|})-L_k(3\frac{\textbf{k}_\bot^2}{|eB|})L_p
(6\frac{\textbf{k}_\bot^2}{|eB|})\right)\right] \nonumber\\&&\frac{k_0p_0-k_3^2-M^2}{(k_0^2-\frac{4}{3}|eB|k-k_3^2-M^2)(p_0^2-\frac{2}{3}|eB|p-p_3^2-M^2)}.
\eeqn
Combing Feynman parameter for the denominator factor and the proper time representation \cite{schwinger1951green}, we obtain \cite{Chao:2014wla}
\be
\frac{1}{ab}=\int_0^1dx\int_0^{\infty}d\tau ~ \tau\exp[(xa+(1-x)b)\tau].
\ee
With the help of generating function of Laguerre polynomials \cite{jeffrey2007table}:
\be
\sum_{t=0}^{\infty}t^nL_{n-i}^a(\xi)=\frac{t^i}{(1-t)^{a+1}}\exp\left[\frac{-t\xi}{1-t}\right],
\ee
for $|t|<1$, we can calculate the summation of Landau level directly.

Performing Taylor expansion at $eB=0$, we set $M_{\rho^-}^2(eB)=M_{\rho}^2(eB=0)-\kappa eB$,
and we obtain $\kappa \approx 2 $ for spin component $s_z=1$ numerically by using the relation
in (\ref{eq:gapequationrho}). The same result can be obtained for $\rho^+$ meson.

\end{document}